\documentclass[a4paper,referee,sn-mathphys,Numbered]{sn-jnl}

\usepackage{graphicx}%
\usepackage{multirow}%
\usepackage{amsmath,amssymb,amsfonts}%
\usepackage{amsthm}%
\usepackage{mathrsfs}%
\usepackage[title]{appendix}%
\usepackage{xcolor}%
\usepackage{textcomp}%
\usepackage{manyfoot}%
\usepackage{booktabs}%
\usepackage{algorithm}%
\usepackage{algorithmicx}%
\usepackage{algpseudocode}%
\usepackage{listings}
\begin{document}

\title{Electron delocalization in a 2D Mott insulator}


\author[1,2]{\fnm{Cosme G.} \sur{Ayani}}

\author[3]{\fnm{Michele} \sur{Pisarra}}

\author[1]{\fnm{Iv\'an M.} \sur{Ibarburu}}

\author[1]{\fnm{Clara} \sur{Rebanal}}

\author[1,4] {\fnm{Manuela} \sur{Garnica}}

\author*[2] {\fnm{Fabi\'an} \sur{Calleja}} \email{fabian.calleja@imdea.org}

\author[2,5] {\fnm{Fernando} \sur{Mart\'{\i}n}}

\author*[1,2,4,6]{\fnm{Amadeo L.} \sur{V\'azquez de Parga}}\email{al.vazquezdeparga@uam.es}

\affil[1]{\orgdiv{Departamento F\'{i}sica de la Materia Condensada}, \orgname{Universidad Aut\'{o}noma de Madrid}, \orgaddress{\street{Cantoblanco}, \city{Madrid}, \postcode{28049}, \country{Spain}}}

\affil[2]{\orgname{IMDEA Nanociencia}, \orgaddress{\street{Calle Faraday 9}, \city{Cantoblanco}, \postcode{28049}, \state{Madrid}, \country{Spain}}}


\affil[3] {\orgdiv{Dipartimento di Fisica}, \orgname{Universit\'a della Calabria and INFN, gruppo collegato di Cosenza}, \orgaddress{\street{Via P. Bucci, cubo 30C}, \city{Rende}, \postcode{87036}, \country{Italy}}}

\affil[4]{\orgname{Instituto Nicol\'as Cabrera}, \orgaddress{\street{Cantoblanco}, \postcode{28049}, \state{Madrid}, \country{Spain}}}

\affil[5]{\orgdiv{Departamento de Qu\'{i}mica, Módulo 13}, \orgname{Universidad Aut\'{o}noma de Madrid}, \orgaddress{\street{Cantoblanco}, \city{Madrid}, \postcode{28049}, \country{Spain}}}

\affil[6]{\orgname{IFIMAC}, \orgaddress{\street{Cantoblanco}, \postcode{28049}, \state{Madrid}, \country{Spain}}}


\abstract{The prominent role of electron-electron interactions in two-dimensional (2D) materials versus three-dimensional (3D) ones is at the origin of the great variety of fermionic correlated states reported in the literature. In this respect, artificial van der Waals heterostructures comprising single layers of highly correlated insulators allow one to explore the effect of the subtle interlayer interaction in the way electrons correlate. In this work, we study the temperature dependence of the electronic properties of a van der Waals heterostructure composed of a single-layer Mott insulator lying on a metallic substrate by performing quasi-particle interference (QPI) maps. We show the emergence of a Fermi contour in the 2D Mott insulator at temperatures below 11K, which we attribute to the delocalization of the Mott electrons associated with the formation of a quantum coherent Kondo lattice. This Kondo lattice introduces a new periodicity in the system, so that the resulting Fermi surface encompasses both the substrate conduction electrons and the now delocalized correlated electrons from the 2D Mott insulator. Density Functional Theory calculations allow us to pinpoint the scattering vectors responsible for the experimentally observed quasi-particle interference maps, thus providing a complete picture of the delocalization of highly correlated electrons in a 2D Mott insulator.}




\maketitle

\section{Introduction}\label{intro}

The transition from a metallic to an insulating state through the continuous variation of external parameters has captivated physicists for the past eight decades. A metal-insulator transition is characterized by the localization of conduction electrons preventing charge transport in the material. Depending on the system, three mechanisms have been proposed to explain this type of transition: the distortion of the crystal \cite{Peierls1955}, the disorder present in the system \cite{Anderson1958} or the correlation between electrons  \cite{Mott1961, Mott1968}.

A lattice model with just one electron per site is expected to be metallic, but N.F. Mott argued that Coulomb repulsion could prevent electrons from jumping from a unit cell to a neighboring one, thus leading to electron localization \cite{Mott1968}. The physics of the Mott insulator is well captured by the Hubbard model, which implies the formation of two sub-bands, one below the Fermi level, occupied by the localized electrons, and another one above the Fermi level, which remains empty. The resulting energy gap between the Hubbard sub-bands reflects the strength of the Coulomb repulsion \cite{Hubbard1963}.

In systems with reduced dimensionality, electronic correlations become more relevant and many-body effects that do not exist or manifest in 3D may prevail. In particular, layered transition metal dichalcogenides (TMDs) represent a family of correlated quasi-2D materials that can exhibit intriguing phenomena such as Ising superconductivity, charge-density-waves, metal-insulator transitions or a quantum spin liquid phase \cite{Wilson1975, Fazekas1979, Sipos2008, Tsen2015, Xi2016, Ruan2021}. Their van der Waals nature offers the possibility to combine different TMDs in a single compound imprinting new features through proximity interactions across interfaces, facilitating the design of artificial structures with unique properties \cite{Geim2013, Novoselov2016, Sierra2021, Guo2023}. For instance, a single layer of 1T-TaS$_{2}$ or 1T-TaSe$_{2}$ has a Mott insulating state related to a Star-of-David (SoD) charge density wave (CDW) \cite{Wilson1975, Fazekas1979, Lee2019}. Interleaving single layers of 1T-TaS$_{2}$ with metallic single layers of 1H-TaS$_{2}$ results in the unconventional superconductivity observed in 4Hb-TaS$_{2}$ crystals \cite{Ribak2020, Nayak2021}. Additionally, it has been suggested that the doping of a 2D Mott insulator, such as copper oxide layers in cuprates, underlies the high-temperature superconductivity observed in those systems \cite{Dagoto1994, Lee2006}. In contrast, the effect of interlayer interactions on the correlated electronic properties in these van der Waals heterostructures remains largely unexplored. 

In this work, we have considered a van der Waals heterostructure consisting of a single layer of 1T-TaS$_{2}$ (a 2D Mott insulator) lying on a crystal of 2H-TaS$_{2}$ (a metallic substrate). From quasi-particle interference (QPI) maps acquired through scanning tunneling microscopy/spectroscopy (STM/STS), we examine the evolution of the electronic properties of the system as a function of temperature. The QPI maps reveal the emergence of a Fermi contour in the 2D Mott insulator when the temperature drops below 11K, indicating the delocalization of the highly correlated Mott electrons. The new Fermi surface encompasses both the conduction electrons of the metallic substrate and the now-delocalized electrons of the 2D Mott insulator, thus showing the importance of interlayer interactions in defining the ground-state electronic properties of the system. With the help of Density Functional Theory (DFT) calculations, we describe the electronic bands of the hybridized 1T/2H system and identify the experimentally detected scattering vectors. 
 
\section{Sample structure}\label{structure}

Bulk 2H-TaS$_{2}$ is metallic and below 78K presents a long-range order quasi-(3$\times$3) incommensurate CDW \cite{Scholz1982, Ayani2022}, associated with the appearance of a pseudo-gap at the Fermi level \cite{Wang1990}.  The 1T-TaS$_{2}$ polymorph below 180K presents a commensurate CDW with an in-plane periodicity of $(\sqrt{13} \times \sqrt{13})\mathrm{R}13.9^{\mathrm{o}}$ dictated by the triangular arrangement of clusters consisting in 13 Ta atoms forming a Star of David (SoD) \cite{Wilson1975, Giambattista1990, Ishiguro1991}. The single occupied 5d orbitals of the 12 outer Ta atoms of the SoD form the valence and conduction bands of the system. The 5d orbital from the remaining Ta atom at the center of the SoD forms a half-filled band. Contrary to the expected metallic behavior, the formation of the CDW is accompanied by a metal-insulator transition attributed to a Mott correlation mechanism \cite{Wilson1975, Fazekas1979}. The presence of the Hubbard gap at the Fermi level has been confirmed by several STM/STS and Angle Resolved Photoemission Spectroscopy measurements at low temperature \cite{Dardel1992, Kim1994, Kim1996, Pillo1999}.  

\begin{figure}[t]%
\centering
\includegraphics[width=1.0\textwidth]{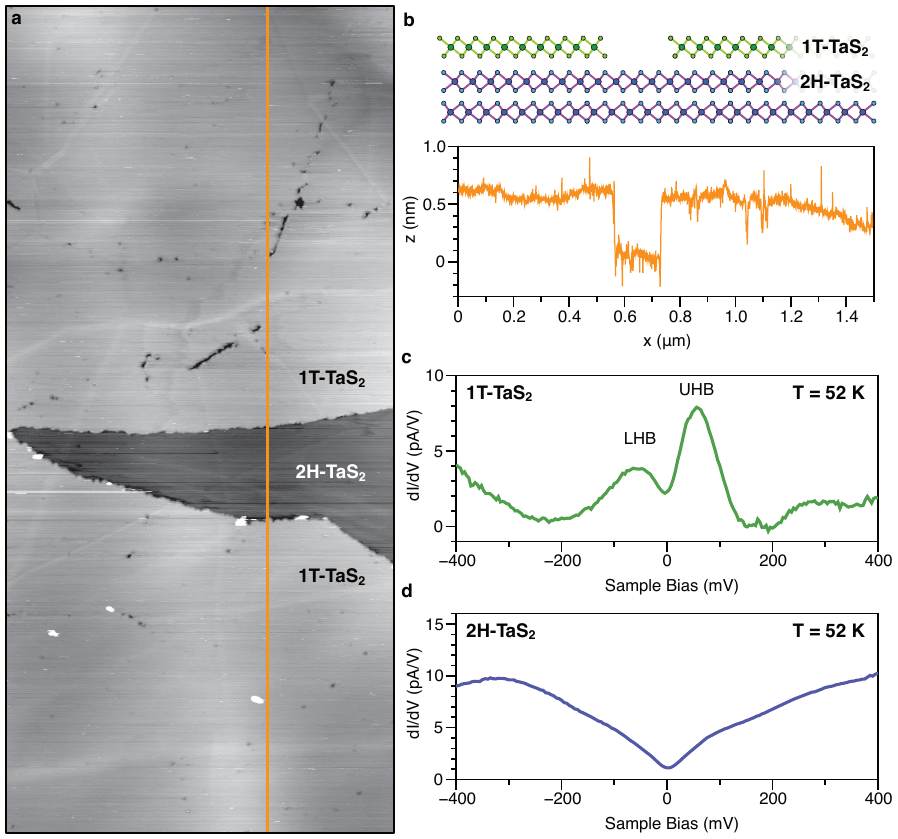}
\caption{\textbf{1T/2H-TaS$_{2}$ sample structure. a,} 1500 nm $\times$ 700 nm section of a large area STM image showing a single layer step-edge. The upper and lower terraces correspond to the 1T and 2H polymorphic phases respectively. Image parameters: V$_{b}$ = 300 mV, I = 100 pA
\textbf{b,} Profile across the single layer step-edge shown in the STM image in panel \textbf{a} with an orange solid line.
\textbf{c,} STS spectrum measured on the 1T terrace showing the presence of the Lower and Upper Hubbard bands (LHB and UHB respectively). STS parameters: V$_{b}$ = 400 mV, I = 500 pA, V$_{mod}$ = 10 mV.
\textbf{d,} STS spectrum measured on the 2H terrace showing its metallic character. STS parameters: V$_{b}$ = 400 mV, I = 500 pA, V$_{mod}$ = 10 mV. All the measurements were performed at 52K.} \label{fig1}
\end{figure}

Figure \ref{fig1}a shows an example of the polymorphic heterostructure comprising a 1T-TaS$_{2}$ single layer on a 2H-TaS$_{2}$ crystal. The V-shaped single layer step edge at the center of the image separates the upper 1T terrace from the lower 2H terrace, where the T and H assignments are based on the CDW identification and STS measurements as explained in the following. The lower panel in Figure \ref{fig1}b displays the line profile corresponding to the orange vertical line in Figure \ref{fig1}a. The apparent step height corresponds to the expected value for a single layer step in a TaS$_{2}$ crystal. The upper panel in Figure \ref{fig1}b is a schematic model of the corresponding structure of the sample. Figure \ref{fig1}c shows a STS spectrum measured at 52K on the 1T terrace, where the Hubbard sub-bands above and below the Fermi level are resolved. Their energy position is renormalized due to the presence of the metallic substrate \cite{Cho2016, Zhu2019a}. Figure \ref{fig1}d displays a STS spectrum measured at 52K on the 2H terrace showing its metallic character and the reduction in LDOS at the Fermi level associated to the quasi-(3$\times$3) CDW \cite{Wang1990}.

\begin{figure}[t]%
\centering
\includegraphics[width=1.0\textwidth]{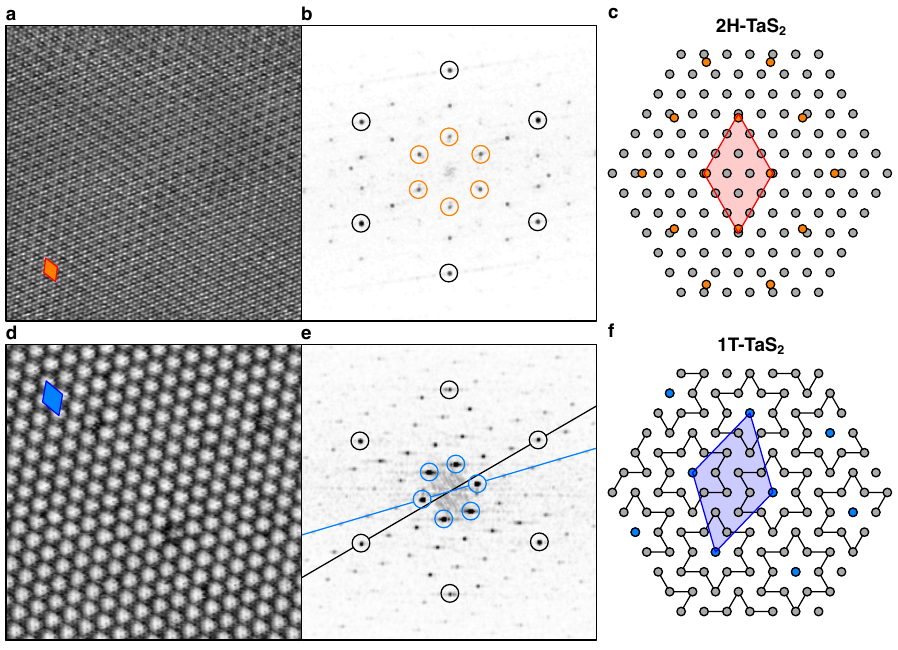}
\caption{\textbf{Charge density waves and atomic structure. a} and \textbf{b}, STM topographic image and corresponding FFT measured on the 2H-TaS$_{2}$ terrace. Both the atomic and the quasi-(3$\times$3) CDW periodicities are resolved and marked with black and orange circles, respectively, in the FFT panel. The CDW unit cell is also marked in orange in the STM panel. Image parameters: 20 nm $\times$ 20 nm, V$_{b}$ = 50 mV, I = 50 pA. \textbf{c,} Ball model of the atomic (grey) and CDW (orange) lattices on the 2H-TaS$_{2}$ terrace. The unit cell of the quasi-(3$\times$3) CDW is highlighted in orange. \textbf{d} and \textbf{e},  STM topographic image and corresponding FFT measured on the 1T-TaS$_{2}$ terrace. Both the atomic periodicity and the $(\sqrt{13} \times \sqrt{13})\mathrm{R}13.9^{\mathrm{o}}$ CDW are resolved and marked with black and blue circles respectively in the FFT panel. The CDW unit cell is also marked in blue in the STM panel. Image parameters: 20 nm $\times$ 20 nm, V$_{b}$ = 500 mV, I = 300 pA. \textbf{f,} Ball model of the atomic (grey) and CDW (blue) lattices of the 1T surface. The black lines mark the SoD clusters. The unit cell of the $(\sqrt{13} \times \sqrt{13})\mathrm{R}13.9^{\mathrm{o}}$ CDW is highlighted in blue.
}
\label{fig2}
\end{figure}

Figure \ref{fig2}a shows an atomically resolved STM image taken on the 2H terrace at 1.2K where both the atomic and the quasi-(3$\times$3) CDW periodicities are apparent, the latter is highlighted with an orange rhombus. Figure \ref{fig2}b shows the Fast Fourier Transform (FFT) of the image shown in panel a. The spots corresponding to the atomic and CDW periodicities are marked with black and orange circles, respectively. In this case, the atomic and CDW lattices are aligned. Figure \ref{fig2}c displays the corresponding real-space ball model, where the atomic and CDW lattices are again represented in grey and orange colors, respectively. The orange rhombus marks the unit cell of the CDW, highlighting its incommensurability. An atomically resolved STM image acquired on the 1T terrace at 1.2K is shown in Figure \ref{fig2}d, both the atomic and the ($\sqrt{13} \times \sqrt{13})\mathrm{R}13.9^{\mathrm{o}}$ CDW periodicities are observed. The blue rhombus marks the unit cell of the CDW. Figure \ref{fig2}e shows the corresponding FFT, where the spots coming from the atomic and CDW periodicities are marked with black and blue circles, respectively. The black and blue lines highlight the $13.9^{\mathrm{o}}$ rotation between the lattices. Figure \ref{fig2}f shows the corresponding real-space ball model where the atomic and CDW lattices are represented in grey and blue colors, respectively. The black lines in this case connect the 12 external Ta atoms of each SoD cluster, as a visual guide. The central Ta atoms as well as the commensurate CDW lattice are highlighted in blue. The superposition of two CDWs with different lattice parameter and orientation leads to the appearance of a moir\'e pattern in the system (see SI section S1).

\section{Delocalization of highly correlated Mott electrons }\label{delocalization}

\begin{figure}[t]%
\centering
\includegraphics[width=1.0\textwidth]{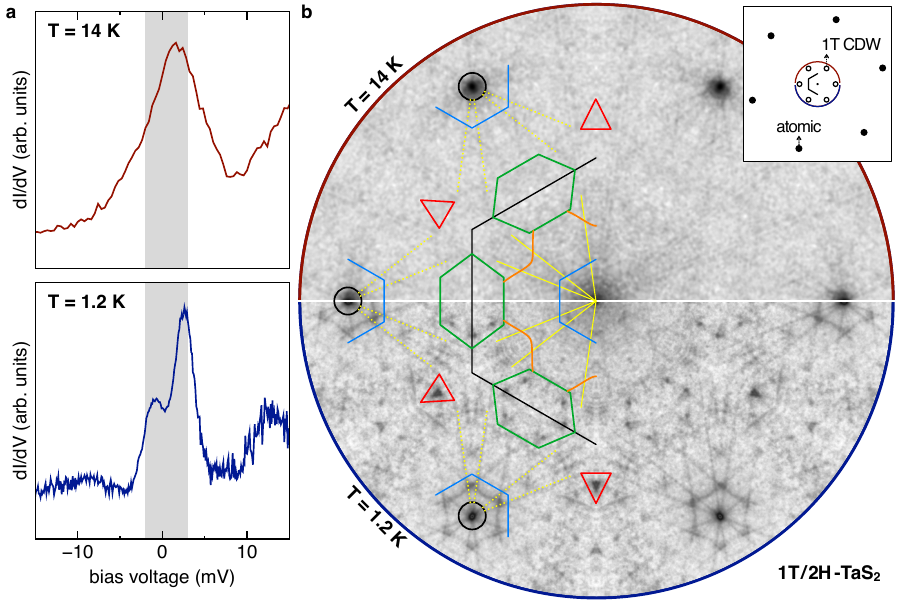}
\caption{\textbf{Tunneling spectroscopy and QPIs. a,} STS spectra recorded on 1T/2H-TaS$_2$ at 14 K (upper panel) and 1.2 K (lower panel). In both cases the Kondo resonance feature is detected (broad peak), while at 1.2 K the Kondo lattice gap appears at the Fermi level superimposed on the Kondo resonance (see region highlighted in grey). \textbf{b,} QPI maps recorded on 1T/2H-TaS$_2$ at 14 K (upper semicircle) and 1.2 K (lower semicircle). The inset in the upper right corner gives a schematic representation of the atomic reciprocal lattice, where the atomic and 1T CDW spots are shown in black filled and empty circles, respectively; the black hexagon represents the 1T CDW SBZ; the blue-red circle marks the region corresponding to the experimental QPI data shown in the main panel. On the left half of the plot, black circles mark the positions of the 1T CDW spots, and the corresponding SBZ is indicated by the black line. Apart from the 1T CDW spots, in the 1.2 K map some additional features are clearly resolved. Outside the 1T CDW SBZ we identify some prominent triangular features, marked as red triangles. Near the SBZ boundaries some groups of spots are encircled by green elongated hexagons. Finally, inside the SBZ we highlight two more features, a flower-like shape marked by the orange curved lines and a central hexagonal structure marked in blue. The latter is replicated around the 1T CDW spots, and marked as well in blue.  Yellow solid lines mark the noise attributed to the scan line direction, as propagated by the symmetrization procedure (see explanation in SI), and the yellow dotted lines are the corresponding replicas emerging from the 1T CDW spots. }
\label{fig3}
\end{figure}

Figure \ref{fig3}a shows the STS spectra measured at the center of the SoD cluster on the 1T-TaS$_{2}$ layer at 14K and 1.2K. The spectrum at 14K, upper panel, shows the presence of a sharp peak at the Fermi level, which is due to the Kondo resonance resulting from the screening of the unpaired electrons localized at the center of the SoD by the conduction electrons of the 2H-TaS$_{2}$ metallic substrate \cite{Vano2021, Ayani2022}. The spectrum measured at 1.2K shows the opening of a narrow gap in the Kondo peak at the Fermi level. The appearance of this gap is the consequence of the coherent quantum superposition of the Kondo clouds associated with the singlets originally located at the center of the SoD clusters \cite{Martin1982}. In this scenario, the electronic structure of both the 1T-TaS$_{2}$ layer and the 2H-TaS$_{2}$ substrate are perturbed by the formation of the coherent Kondo lattice, with two main consequences. The first one is that, since the Kondo resonances become electronically periodic, Bloch's theorem ensures the appearance of a band at the Fermi level of width of the order of the energy scale associated to the Kondo screening of the individual magnetic moments, indicated by the Kondo temperature, T$_{\mathrm{K}}$. This band hybridizes with the metallic bands of the substrate and a gap opens in the middle of the Kondo resonance, whose width is dictated by the strength of the hybridization \cite{Martin1982}. The second consequence is the de-localization of the Mott electrons, that should become part of the Fermi surface of the new periodic crystal \cite{Martin1982, Hewson1993}. 

We have performed a detailed analysis of the modifications of the Fermi surface resulting from the formation of the Kondo lattice by obtaining quasi-particle interference maps from STM measurements. It is well known that defects present on surfaces act as scattering centers for the conduction electrons leading to interferences (hence the term QPI) that result in the formation of standing waves in the surface LDOS \cite{Crommie1993}. Therefore, the presence of a QPI pattern at the Fermi level implies the existence of conduction electrons and, therefore, of a Fermi contour. Furthermore, the Fermi contour itself is ``encoded'' into the QPI pattern through the scattering selection rules, defect distribution and defect shape among other factors \cite{Inoue2016,Russman2021,Marques2021}.

We acquired dI/dV curves in every pixel of several STM images covering extensive regions of the 1T/2H-TaS$_{2}$ heterostructure, sweeping the bias voltage across the Fermi level (see SI section S2 for details). The resulting dI/dV maps were averaged, Fourier transformed and symmetrized following the procedure described in the SI, section S3. Figure \ref{fig3}b shows two of the resulting QPI maps recorded at 14K (upper half) and 1.2K (lower half), integrating an energy window from -2.5mV to +3.0mV around the Fermi energy (grey area in Figure \ref{fig3}a). The corresponding 1T surface Brillouin zone (SBZ) is marked as a black hexagon (only half of it, for clarity). Note that these maps cover a relatively small region of the reciprocal space in comparison to that associated with the atomic lattice, as illustrated in the inset in Figure \ref{fig3}b.

In the QPI map measured at 14K, apart from the 1T CDW spots, the only detected signals are the noise lines related to the scan line direction, always kept around 8$^{\mathrm{o}}$ with respect to a high symmetry direction of the CDW. These noise lines, highlighted in yellow, are mirrored and propagated due to the symmetrization procedure followed to improve the signal to noise ratio (see SI section S3). The yellow dotted lines are the corresponding replicas emerging from the CDW spots. This lack of QPI signal at the Fermi level at 14K is consistent with the Mott insulating state of the 1T layer above the coherence temperature for the Kondo lattice. On the contrary, the QPI map measured at 1.2K (lower half of Figure \ref{fig3}b) shows much more structure than its 14K counterpart. In this case, apart from the 1T CDW spots (marked with black circles) and the experimental noise already explained in the 14K map (and marked with yellow lines), a number of new spots and structures can be clearly resolved. In particular, around the $\Gamma$ point we resolve a hexagonal structure, marked in blue. This structure replicates around the 1T CDW spots, where it is even better resolved (also marked in blue). Moving away from the $\Gamma$ point, we detect a "flower-like" structure, marked in orange. Near the SBZ edge, there is a group of six spots that we have surrounded with green hexagons. Finally, outside the 1T CDW SBZ there are intense triangular spots, each one marked with a red triangle. Note that none of these structures are present in the 14K map.  

\begin{figure}[t]%
\centering
\includegraphics[width=0.93\textwidth]{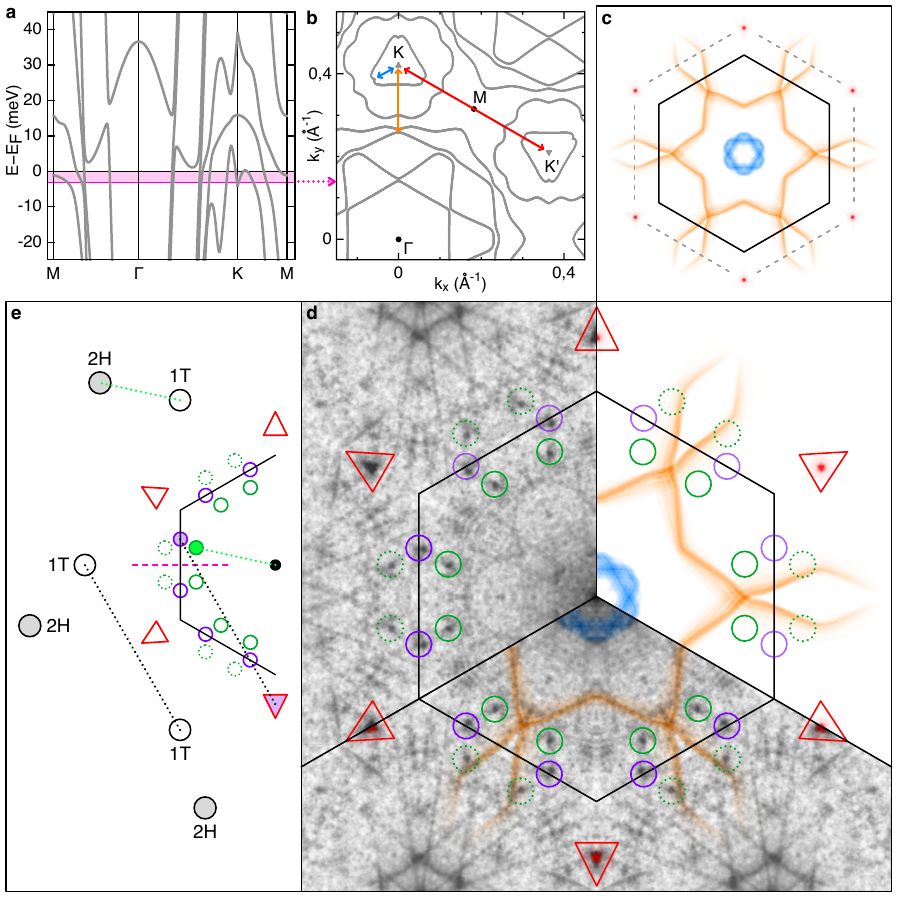}
\caption{\textbf{Tunneling spectroscopy and QPI maps. a,} DFT calculated energy bands of the hybridized system. \textbf{b,} Constant energy contour at -3 meV calculated from the DFT band structure shown in panel a. The symmetry points of the 2H SBZ are indicated by the black letters, and colored arrows (blue, orange and red) indicate the scattering processes considered for the QPI simulations shown in panels c and d, integrating the energy window from the Fermi level to -3meV (cfr. the pink shaded region in panel a). \textbf{c,} QPI simulation of the scattering processes indicated by arrows in panel b, using the same color code. The 2H and 1T SBZs are indicated by the grey dashed and black solid lines, respectively. \textbf{d,} QPI simulation overlaid on the experimental data recorded at 1.2 K. The panel is divided into three 120$^\circ$ sections. The upper right section shows the simulated QPI shown in panel c. The upper left section shows the experimental QPI map shown in the lower half of Figure \ref{fig3}b. The lower section shows the superposition of the simulated and experimental QPI maps with the 1T SBZ indicated by the black hexagon. The red triangles mark the points corresponding to the K-K' inter-valley process indicated by the red arrow in panel b. Inside the SBZ the flower-like structure is due to the inter-valley scattering marked with an orange arrow in panel b. The hexagonal structure around the $\Gamma$ point is due to the inter-valley scattering marked with a blue double arrow in panel b. The purple circles at the edge of the SBZ mark the replicas of the red triangles due to the 1T CDW periodicity, as explained in panel e. Green solid circles inside the SBZ mark the periodicity of the moir\'{e} pattern due to the superposition of the 2H and 1T CDWs, as explained in panel e. The green dotted counterparts are the replicas outside the SBZ. \textbf{e,} Model of the reciprocal space of the system. Grey-filled and hollow-black circles represent the first order spots of the 2H and 1T CDWs, respectively. The black half hexagon marks the 1T SBZ. The green dotted line corresponds to the lattice vector of the moir\'{e} resulting from the superposition of the 2H and 1T CDWs. The green circles indicates the moir\'{e} superperiodicity. The black dotted line corresponds to the lattice vector of the 1T CDW and projects the red triangular features on the edge of the SBZ marked as solid purple circles.}
\label{fig4}
\end{figure}

We trace back the origin of the additional signals detected in the QPI maps measured at 1.2K with the help of band structure calculations. The 2H-TaS$_2$ bulk electronic structure is obtained by means of DFT calculations, taking into account the formation of the CDW periodicity (see SI section S4). The formation of the coherent Kondo lattice is described by introducing a quasi-flat band at the Fermi level with the periodicity of the system and a width of the order of T$_\mathrm{K}$ \cite{Martin1982}. This band is coupled to the conduction bands of the metallic substrate, which results in the opening of gaps at the Fermi level (see section S4 in the SI) \citep{Ayani2022, Ayani2022a}. The resulting band structure is shown in Figure \ref{fig4}a, and the corresponding constant energy contour (CEC) at -3 meV is presented in Figure \ref{fig4}b. Figure \ref{fig4}c shows the band selective QPI simulations for the scattering vectors indicated in Figure \ref{fig4}b by the red, orange and blue double arrows, integrating the energy window from the Fermi level to -3meV, as indicated by the pink shaded region in Figure \ref{fig4}a (see SI section S5). The simulations are overlaid on the 1.2K experimental data in Figure \ref{fig4}d, which has been divided into three 120$^\circ$ sections. The upper right section shows the same data as Figure \ref{fig4}c, the QPI simulation for the scattering processes indicated in Figure \ref{fig4}b, using the same color code. The upper left section shows the experimental data shown in Figure \ref{fig3}b, in grey color code. The lower section is the combination of both. The 1T SBZ is indicated by the black solid hexagon throughout the three segments.

The hexagon surrounding the $\Gamma$ point at the center of the SBZ in Figure \ref{fig4}d is the result of the inter-valley scattering between the bands around the K point, marked with a blue arrow in Figure \ref{fig4}b. The simulated QPI signal is superimposed in blue. Moving outwards from the center of the SBZ, the flower-like structure is the result of the scattering between the bands at the K point and bands around the $\Gamma$ point, marked with an orange double arrow in Figure \ref{fig4}b. Moving towards the border of the 1T SBZ, the 6 spots, indicated by green elongated hexagons in Figure \ref{fig3}b, are marked here in Figure \ref{fig4}d by purple and green circles. The origin of the spots located inside and outside the SBZ (green solid and dotted circles, respectively) is the moir\'e pattern arising from the superposition of the 1T and 2H CDW periodicities (see section S1 in SI). This is explained in Figure \ref{fig4}e where some of the 2H and 1T CDW spots are represented in grey-filled and hollow circles, respectively. Projecting from the origin, the difference vector between the 2H and 1T CDWs (green dotted line), which by definition is the moir\'e vector, results in the spot marked as a green filled circle inside the 1T SBZ. Due to the symmetrization procedure (see SI section S3), this spot will be mirrored along the pink dashed line and replicated over the different high symmetry directions, producing the set of solid green hollow circles. Finally, all these spots should be mirrored also across the SBZ boundaries, leading to the dotted green circles. All these circles match perfectly the experimental spots, whose origin can thus be attributed to the periodicity of the moir\'e pattern resulting from the superposition of both CDWs. The fact that these spots are present only in the 1.2K map and not in the 14K one (see Figure \ref{fig3}b) indicates that at 1.2K we are sensitive to both the 1T and 2H periodicities, despite the fact that we are tunneling into the 1T overlayer only. This is understood as a consequence of the already mentioned hybridization between the 1T flat band describing the periodic Kondo lattice and the conduction bands of the 2H substrate, resulting in the de-localization of the 1T Mott electrons into the 2H band structure. Hence, the electronic structure of the combined system  incorporates the symmetry of both the 1T and 2H CDWs.

Finally, the spots marked with red triangles located outside the 1T SBZ match the size of the SBZ of the CDW of the 2H layer. In fact, the simulation for the K-K' inter-valley scattering indicated by the red double arrow in Figure \ref{fig4}b leads to a very localized signal precisely at these locations. The spots located at the 1T SBZ border (purple circles) can be explained as replica spots from the ones marked with red triangles. This is explained in Figure \ref{fig4}e, where the black dotted line corresponds to a lattice vector from the 1T CDW, as it connects two of its consecutive spots (black circles). The projection of the same lattice vector from the position of one of the triangular features ends up at the border of the 1T SBZ, precisely at the position of one of the features detected in the experiment (purple filled circle). The fact that we experimentally detect the K-K' inter-valley scattering replicas associated with the 1T CDW periodicity (purple circles in Figure \ref{fig4} d and e) confirms that we are tunneling into the 1T layer while having access to the electronic structure of the 2H substrate. Taking into account the symmetrization procedure already mentioned, this spot will be mirrored and propagated, resulting in the positions given by the purple hollow circles. Note that neither the CDW moir\'e spots (green circles in Figure \ref{fig4}d) nor the replica spots from the main K - K' inter-valley process (purple circles in Figure \ref{fig4}d) are present in the simulation shown in Figure \ref{fig4}c, as expected because the simulation does not include the 1T CDW periodicity. 

In conclusion, as is well known, a single layer of 1T-TaS$_{2}$ goes through a metal-insulator transition at 180K. Here we have shown that it becomes metallic again below 11K when lying on 2H-TaS$_{2}$. The transition back to a metallic state as temperature goes down is attributed to the formation of a coherent Kondo lattice, thus introducing a new periodicity, which, according to Bloch's theorem, results in the appearance of a flat band at the Fermi level with a width of the Kondo temperature. This band ultimately hybridizes with the metallic bands of the 2H-TaS$_{2}$ substrate, so that the new Fermi surface of the system incorporates both the conduction electrons from the latter and the now-delocalized electrons from the 1T-TaS$_{2}$ Mott insulator, in accordance with  Luttinger's sum rule. By measuring QPI maps, we have observed the formation of the Fermi contour in the 1T-TaS$_{2}$ layer below 11K and, with the help of DFT calculations, identified the origin of the prominent features in those maps. In the end, we have shown how the interlayer interaction in a van der Waals heterostructure allows for the reversal of the metal-insulator transition while preserving the CDW of the system. As a side note, the appearance of a Fermi contour below 11K solves the controversy on the nature of the ground state in similar van der Waals heterostructures, which has been attributed to a variety of effects, such the formation of heavy-fermions, a magnetically ordered ground state or a doped Mott insulator \cite{Vano2021, Wan2022, Crippa2023}. Thus, our results provide additional and fundamental insight on the importance of interactions between layers in van der Waals heterostructures and pave the way to induce novel properties in highly correlated 2D systems.

\section{Methods}

\bmhead{Sample preparation}

The 2H-TaS$_{2}$ single crystal was exfoliated at room temperature in UHV using a Nitto tape and then it was transferred to the Low Temperature STM without breaking the UHV conditions.

\bmhead{STM/STS measurements and tip preparation}

The STM experiments were carried out in a low-temperature STM operating in Ultra-High Vacuum at a base pressure in the low $10^{-10}$ mbar. The UHV system is equipped with a preparation chamber and a load-lock that allows sample exfoliation and transfer to the STM without breaking the UHV conditions at any point. Two different temperature regimes were employed during the measurements, one with the STM cooled with liquid nitrogen and pumped down with a scroll pump to lower the base temperature to 52K, and a second one with the STM cooled to a base temperature of 1.2 K. In this second regime two different stable temperatures were used in order to perform the (several days long) QPI maps, the already mentioned 1.2 K base temperature of the cryostat, and an intermediate temperature of 14 K. The latter was achieved by leaving open the two inner radiation shields of the cryostat (helium bath and Joule-Thomson pot) and, at the same time, applying heat with an internal resistive heater located at the JT pot and controlled by a PID circuit. This configuration allowed for a stable enough temperature in order to perform the QPI maps as described in section S2 of the SI.

The STS spectra have been measured using a lock-in amplifier, adding a modulation to the sample bias voltage signal between 4mV and 500 $\mu$V, depending on the particular experiment. In all cases the modulation frequency was 763 Hz.

The STM tips were home-made by electrochemical etching of a tungsten wire and cleaned in UHV by argon sputtering. Before the experiments the tips were checked against a Cu(111) sample to ensure that they presented a flat density of states at the Fermi level while resolving the well-known Cu(111) surface state without any additional features. An Au ball made from a 99.99$\%$ pure wire was adhered to the sample holder close to the TaS$_{2}$ crystal so that the tip could be prepared \textit{in-situ} by performing indentations in the Au ball without opening the STM thermal shields and therefore keeping the sample thermalized and clean. 

\bmhead{DFT modelling}

DFT calculations for the 2H phase of the bulk TaS$_{2}$ crystal have been carried out within the Projector Augmented Wave (PAW) method \cite{Blochl1994} as implemented in the Vienna Ab Initio Simulation Package (VASP) \cite{Kresse1996, Kresse1996a, Kresse1999}. We used a 350 eV plane wave cut off, the Perdew Burke Ernzerhof (PBE) functional \cite{Perdew1996} and the Grimme D3 correction \cite{Grimme-D3}, which proved to be adequate for 2H transiton metal dichalcogenides \cite{Pisarra-TMD}. We adopted a very strict 10$^{-6}$ eV convergence criterion for the self-consistent electronic cycles and $(18\times18\times4)$ and $(6\times6\times4)$ unshifted Monkhorst-Pack grids for the Brillouin Zone (BZ) sampling of the $(1\times1)$ and $(3\times3)$ 2H-TaS$_2$ reconstruction, respectively. All the geometry optimizations have been carried out optimizing the position of all the atoms until the maximum residual force was lower than 10$^{-3}$ eV/\AA. 

The Kondo lattice formation was modelled using, as conduction band, the DFT calculated $k_z=0$ portion of the band structure of the 2H-TaS$_2$ system in the $(3\times3)$ CDW reconstruction. The impurity band, on the other hand, had the typical dispersion for TaS$_2$, that is a global maximum at the $\Gamma$ point, local maxima at the K points of the BZ, saddle points at the M points and minima along the $\Gamma$M direction, and a very reduced bandwidth of $\sim40$~meV. The conduction and impurity band have been coupled through the effective Hamiltonian~\cite{Martin1982}:
\begin{equation}
\nonumber H{=} \sum_{\mathbf{k}n} \epsilon_{\mathbf{k}n} d^\dagger_{\mathbf{k}n} d_{\mathbf{k}n} + \sum_{\mathbf{k}} \varepsilon_{\mathbf{k}} c^\dagger_{\mathbf{k}} c_{\mathbf{k}} + V (\sum_{\mathbf{k}n}  c^\dagger_{\mathbf{k}} d_{\mathbf{k}n} + h.c.)
\end{equation}
where $\epsilon_{\mathbf{k}n}$ is the one electron energy of the conduction band (generated by the creation and annihilation operators $d^\dagger_{\mathbf{k}n}$, $d_{\mathbf{k}n}$), $\varepsilon_{\mathbf{k}}$ is the impurity band (generated by the creation and annihilation operators $c^\dagger_{\mathbf{k}}$, $c_{\mathbf{k}}$), $n$ is the conduction band index, $\mathbf{k}$ is sorted within the 1$^{\mathrm{st}}$BZ, and the coupling potential $V=5$~meV is assumed to be $\mathbf{k}$- and band-independent~\cite{Ayani2022}. More details about the DFT modelling of the Kondo Lattice formation are available in section 4 of the SI.

\backmatter

\bmhead{Supplementary information}

Supplementary information is available for this paper.

\bmhead{Acknowledgments}

This work was supported by Ministerio de Ciencia, Innovaci\'on y Universidades (MICIU/AEI/10.13039/501100011033) through grants, PID2021-128011NB-I00, PID2022-138288NB-C31 and “Ayudas para Incentivar la
Consolidaci\'on Investigadora” (CNS2022-135175), Comunidad de Madrid through grants "Materiales Disruptivos Bidimensionales (2D)" (MAD2D-CM)-UAM and MAD2D-CM-IMDEA-NC funded by the Recovery, Transformation and Resilience Plan, and by NextGenerationEU from the European Union. IMDEA Nanoscience acknowledges support from the "Severo Ochoa" Programme for Centres of Excellence in R\&D CEX2020-001039-S. IFIMAC acknowledges support from the "Mar\'{\i}a de Maeztu" Programme for Units of Excellence in R\&D CEX2018-000805-M. MG thanks Ministerio de Ciencia, Innovaci\'on y Universidades "Ram\'on y Cajal" Fellowship RYC2020-029317-I. MP acknowledges financial support by “Centro Nazionale di Ricerca in High-Performance Computing, Big Data and Quantum Computing”, PNRR 4 2 1.4, CI CN00000013, CUP H23C22000360005. We acknowledge allocation of computing time at the Centro de Computaci\'on Cient\'{\i}ifica at the Universidad Aut\'onoma de Madrid, the CINECA Consortium INF16-npqcd Project and Newton HPCC Computing Facility at the University of Calabria (MP).

\section*{Declarations}

The authors declare no conflict of interest. 


\bibliography{QPI_refs}

\end{document}